\documentstyle[12pt,fleqn]{article}
\def\singlespace {\smallskipamount=3.75pt plus1pt minus1pt
                  \medskipamount=7.5pt plus2pt minus2pt
                  \bigskipamount=15pt plus4pt minus4pt
                  \normalbaselineskip=15pt plus0pt minus0pt
                  \normallineskip=1pt
                  \normallineskiplimit=0pt
                  \jot=3.75pt
                  {\def\smallskip {\vskip\smallskipamount}}
                  {\def\medskip   {\vskip\medskipamount}}
                  {\def\bigskip   {\vskip\bigskipamount}}
                  {\setbox\strutbox=\hbox{\vrule 
                    height10.5pt depth4.5pt width 0pt}}
                  \parskip 7.5pt
                  \normalbaselines}
\def\middlespace {\smallskipamount=5.625pt plus1.5pt minus1.5pt
                  \medskipamount=11.25pt plus3pt minus3pt
                  \bigskipamount=22.5pt plus6pt minus6pt
                  \normalbaselineskip=22.5pt plus0pt minus0pt
                  \normallineskip=1pt
                  \normallineskiplimit=0pt
                  \jot=5.625pt
                  {\def\smallskip {\vskip\smallskipamount}}
                  {\def\medskip   {\vskip\medskipamount}}
                  {\def\bigskip   {\vskip\bigskipamount}}
                  {\setbox\strutbox=\hbox{\vrule 
                    height15.75pt depth6.75pt width 0pt}}
                  \parskip 11.25pt
                  \normalbaselines}
\def\doublespace {\smallskipamount=7.5pt plus2pt minus2pt
                  \medskipamount=15pt plus4pt minus4pt
                  \bigskipamount=30pt plus8pt minus8pt
                  \normalbaselineskip=30pt plus0pt minus0pt
                  \normallineskip=2pt
                  \normallineskiplimit=0pt
                  \jot=7.5pt
                  {\def\smallskip {\vskip\smallskipamount}}
                  {\def\medskip   {\vskip\medskipamount}}
                  {\def\bigskip   {\vskip\bigskipamount}}
                  {\setbox\strutbox=\hbox{\vrule 
                    height21.0pt depth9.0pt width 0pt}}
                  \parskip 15.0pt
                  \normalbaselines}

\include{dspace12}
\include{epsf}
\def\be{\begin{equation}}
\def\ee{\end{equation}}
\def\bea{\begin{eqnarray}}
\def\eea{\end{eqnarray}}

\def\sect #1{\setcounter{equation}{0}}

\begin{document}
\singlespace
\begin{center}
\large{\bf Reissner Nordstr\"{o}m Background Metric in Dynamical Co-ordinates:
       Exceptional Behaviour of Hadamard States}
\end{center}

\begin{center}
{Sukratu  Barve}\footnote{sukratu@prl.res.in 
  Phone: 091 079 630 2129 x 4775}\\
{A.R.Prasanna}\footnote{prasanna@prl.res.in 
 Phone: 091 079 630 2129 x 4475}\\
   Theoretical Physics Division\\
   Physical Research Laboratory\\
   Navrangpura\\
   Ahmedabad 380 009\\
   INDIA\\ 
\end{center}

\abstract
          { We cast the Reissner Nordstrom solution in a particular 
co-ordinate  system which shows dynamical evolution from initial data. 
The initial data for the $E<M$ case is regular. This procedure enables 
us to treat the metric as a collapse to a singularity. It also implies 
that one may assume Wald axioms to be valid globally in the Cauchy 
development, especially when Hadamard states are chosen. We can thus 
compare the semiclassical behaviour with spherical dust case, looking 
upon the metric as well as state specific information as evolution from 
initial data.  
             We first recover the divergence on the Cauchy horizon 
obtained earlier. We point out that the semiclassical domain extends 
right upto the Cauchy horizon. This is different from the spherical dust 
case where the quantum gravity domain sets in before. We also find that 
the backreaction is not negligible near the central singularity, unlike 
the dust case.
             Apart from these differences, the Reissner Nordstrom solution 
has a similarity with dust in that it is stable over a considerable period 
of time. 

The features appearing dust collapse mentioned above were suggested to be 
generally applicable within spherical symmetry. Reissner Nordstrom background 
(along with the quantum state) generated from initial data, is shown not to 
reproduce them.}
\newpage

\section{Introduction}
	The issue of Cosmic Censorship has several physical implications, two of which have been in focus recently. 

	Firstly, it appears that spacetime singularities might signal the 
visibility of high curvature regions.
 Even if completely absent in loop quantum gravity \cite{bojowald}, classical 
singularities would 
correspond to regions of high curvature and nakedness would imply their 
exposure to asymptotic observers. This itself could lead to observational 
signatures of particle creation and such effects \cite{tpplanck}.

	Secondly, there appears to be a contrast in the semiclassical behaviour
 of black holes and naked singularities. That contrast could be carried 
through into the quantum gravity domain. This has been clearly brought out in 
the dust collapse case \cite{tpplanck}. The covered singularity evaporates 
leaving behind
 about a Planck mass before reaching the quantum gravity domain. However, when
 the initial data corresponds classically to naked singularities,
most of the star collapses almost intact into the quantum gravity domain
all through the semiclassical phase . The star stays almost classical directly
till the quantum gravity stage. Moreover, this happens well within the 
lifetime of the Universe \cite{tpplanck}. The naked singular solution is thus 
special from the physical point of view.  This argument is subject to the
 assumption that the behaviour in the dust case is typical of the general 
scenario. 
        
         Out of these, we focus on the second aspect in this paper. Although
 a contrast in behaviour may still occur, the analysis of dust collapse does 
not warrant to be considered as typical as suggested in \cite{tpplanck}. The 
behaviour of dust depends not only on the semiclassical phase but also on the 
onset of the quantum gravity domain as well as backreaction considerations. In
 the spherical dust case, the quantum gravity domain is reached at some stage 
before the formation of the Cauchy horizon \cite{tpplanck} and the 
backreaction is found to be negligible. 
 
The behaviour above need not always be the case. To this end, we show that the
 Reissner Nordstrom  metric (charge denoted by $E$ and mass by $M$) $E<M$ case is an exception.  
Before proceeding on that, we first point out that the 
Reissner Nordstr\"{o}m metric can be presented in co-ordinates which mimic 
dynamical evolution. As far as we are aware, this has not been done previously.
 This leads us to two important technical facts. Firstly, one can work with
 Hadamard states, in principle requiring specification only on the initial Cauchy surface. This justifies the global validitity of Wald axioms for the quantum
 stress tensor which will be assumed in this work. Secondly, we can easily imagine the metric to result from initial data since we have explicit dynamical 
co-ordinates. This qualifies the metric to be considered as a collapse leading to a singularity (curvature singularity).
  The case $E<M$ can be obtained from non-singular initial data. We then 
calculate the quantum stress tensor for massless scalar fields
 on this background and recover the divergence of the quantum stress tensor
 on the Cauchy horizon obtained earlier \cite{hiscock}. It should be noted that
 the global validity of Wald axioms is essential in this and similar results
 concerning the quantum stress tensor. This has not been justified in previous work on Reissner Nordstrom metric.

It should be noted that the Reissner Nordstr\"{o}m solution is likely to
be more of an idealized situation. A realistic collapse would perhaps 
be accompanied by a radiative influx.
 Perturbations of the solution are therefore extensively 
studied especially for instability of Cauchy horizon \cite{hiscockrn} 
\cite{penrose} \cite{brady} \cite{poisrael}. Several issues arise therefrom. 
Some, like the inflation of the mass function, suggest a non local characterization of the Cauchy horizon in special models. Others admit observer dependence of divergences. 
The premise for our theme necessitates consideration of only local quantities, the divergences of which are observer independent. We wish to address the issues in the simplest of cases and present the Reissner Nordstr\"{o}m solution as one.  

 Though 
we consider the evolution for initial data corresponding to the Reissner 
Nordstr\"{o}m solution, the result is useful in so far as concerns of a
general possibility of an energy burst near Cauchy horizons. In general, one
needs to account for backreaction and invoke quantum gravity to check this.
In this example, the $E<M$ initial configuration
radiates away right upto the Cauchy horizon. The quantum gravity limit does 
not occur before the horizon. This is unlike the dust case scenario (suggested 
to be generally valid within spherical symmetry) wherein the quantum gravity domain sets in earlier.\\ 
We thus wish to point out that the interplay between physical limits needs to 
be considered carefully, even within
 general spherical symmetry. The counterexample of initial data that appears
as this simple case, warrants further investigation into the issue. 

Apart from this theme , we note that our result for our example does not 
strengthen the case for a burst because of backreaction.
Also, there is a similarity with dust collapse in our example. We find
that the configuration does stay intact till it is very close to the 
Cauchy horizon.

\section{Quantum Stress Tensor and choice of state}
	Much of the physics of quantum fields in curved spacetime is captured by a local quantity, the quantum stress tensor. See for instance the chapter 6
 of \cite{BD} (The central problem). After several approaches to regularize and renormalize this quantity, R. M. Wald came up with four axioms which restricted the form of the quantum stress tensor to a very large extent \cite{waldqft}.
 
From Wald axioms and the trace anomaly in 2-d, one can show that
 the quantum stress tensor is of the form 

\bea
\label{genq}
<T_{uu}>&=&C\left(1/C\right)_{,u,u}+ {\cal AA}(u)\\
<T_{vv}>&=&C\left(1/C\right)_{,v,v}+ {\cal BB}(v)\\
<T_{uv}>&=&{\cal R}/24 \pi
\eea

where the line element is expressed in double null co-ordinates as 
$ds^{2}= C^{2}dudv$ and ${\cal R}$ is the scalar curvature of the background 
spacetime.
The information about the state is contained in ${\cal AA}$ and ${\cal BB}$.
It can be easily argued based on asymptotic behaviour of $<T_{\mu\nu}>$ that
${\cal AA}$ and ${\cal BB}$ are finite in their domain of definition \cite{hiscock}.

	The important assumption here is that the Wald axioms, especially the conservation of the stress tensor, hold all over the spacetime. Physically, one 
would like to ensure that only by specifying initial conditions near a Cauchy 
surface, expecting the equations to hold all through the evolution. Indeed 
this is possible if one chooses Hadamard states. For a definition 
 and related mathematical development see \cite{hadamard} and \cite{kaywald}. 
These states can be defined using the singularity structure of the two point 
function. They have two useful properties a) that they satisfy Wald axioms if 
they stay Hadamard and b) by choosing the singularity structure near a Cauchy 
surface, one is essentially guaranteed that the Hadamard form is preserved all 
throughout the Cauchy development \cite{FSW}.  

	In our example, the choice of state is automatically contained in the choice of the ${\cal AA}$ and ${\cal BB}$. Certainly, there would be choices 
corresponding to Hadamard states and that poses no conceptual difficulty.
 However, property b) requires careful attention since it would restrict 
considerations to Cauchy developments only. In particular, if the spacetime or 
region of spacetime under consideration is realized as a Cauchy development then we can assume properties a) and b), and thereby we can be assured of the formulae
equations (\ref{genq}) to hold all over the region.

	We show that the Reissner Nordstr\"{o}m solution can be explicitly 
realized as a time evolution of initial data justifying our use of Wald axioms
 in a global sense. In fact we show that we can re-cast the metric in co-ordinates such that the evolution becomes explicit. 

\section{Reissner Nordstr\"{o}m solution using special co-ordinates}
We develop the first part of the analysis in this section, wherein we  
re-cast the general Reissner Nordstr\"{o}m solution. 
However, for further analysis, only $E<M$ is suitable. 
 
We consider the field equations for spherical symmetry. In the first subsection,  we make certain assumptions on the radial pressure and on the 
gravitational potential. We show in the next subsection, that further choice 
recovers the Reissner Nordstr\"{o}m metric. This enables us to cast the metric
 in a form similar to dynamical evolution.
This has been done for the Schwarzschild metric \cite{LL}.
In the following subsection we show that the condition for nakedness is indeed $E^{2}>M^{2}$. Finally we present the Penrose diagram illustrating the region covered by the co-ordinates.

\subsection{Spherical symmetry}

The line element in co-ordinates $(t,r,\theta,\phi)$ is given by
\be
\label{sphline}ds^{2}=e^{\sigma(t,r)}dt^{2}-e^{\omega(t,r)}dr^{2}-R^{2}(t,r)d\Omega^{2}
\ee
We use a source 
\be 
\label{source}{\large T}^{\mu}_{\nu}= diag[ \rho(t,r), -p_{r}(t,r), -p_{T}(t,r), -p_{T}(t,r)]
\ee
 
The general field equations for spherical symmetry [See for example \cite{tang}and references therein] are  

\be
\label{m'}  2m'(t,r)=8\pi\rho (t,r) R^{2}R'
\ee

\be
  \dot{2m}(t,r)=-8\pi p_{r}(t,r)R^{2}\dot{R}
\ee

\be 
  \sigma'(t,r)=-\frac{2\rho'}{\rho+p_{r}}+\frac{4R'(p_{T}-p_{r})}{R(\rho+p_{r})}
\ee

\be
  \dot{\omega}(t,r)=-\frac{2\dot{\rho}}{\rho+p_{r}}-\frac{4\dot{R}(\rho + p_{T})}{R(\rho+p_{r})}
\ee

 where 

\be
 2m(t,r)/R-1=e^{-\sigma}\dot{R}^{2}-e^{-\omega}R'^{2}
\ee

Here, $m(t,r)$ is a free function arising out of integration of the Einstein equations. Its initial value $m(0,r)$ is interpreted as the mass to the interior of the shell with co-ordinate $r$.

In order to proceed towards obtaining the Reissner Nordstr\"{o}m metric, we impose  certain assumptions within spherical symmetry. These are similar to but more general than those in \cite{tang}. We cast them in a particular form so that the generalization is evident.
 
 \be
 8\pi p_{r}R^{2}= {\cal A}(r) {\cal B'}(R)
\ee

and

\be
 \sigma'(t,r)= \psi(r)Q'(R)R'
\ee

 From the last assumption above using the fact that the Ricci tensor component ${\cal R}_{tr}=0$, we obtain

 \be
\label{I} R'^{2}e^{-\omega}=A(r)e^{Q(R)\psi(r)} 
\ee  

where $A(r)$ is arbitrary.

From the time rate of change of $2m$, and with the second last assumption, we 
get  

\be
\label{fin2m}2m(t,r)= {\cal A}(r){\cal B}(R)+2m_{o}(r)
\ee

where $2m_{o}(r)$ is arbitrary and ${\cal B}(R)=\int {\cal B'}(R)dR$

Further, from the definition of $2m$, we obtain, using these results,

\be
\label{III}\dot{R}^{2}e^{-\sigma}=\frac{{\cal A}(r){\cal B}(R)+2m_{o}(r)}{R}-1+A(r)e^{Q(R)\psi(r)} 
\ee
 This is the evolution of $R$ in time. Given $2m_{o}(r)$, and using equation(\ref{fin2m}), we can obtain the 3-metric (and also extrinsic curvature) at any time. 
 Thus we have set up a system of evolution for initial data, much like that in
 \cite{tang}.

\subsection{Reissner Nordstr\"{o}m Solution}

 We now show that the Reissner Nordstr\"{o}m metric can be cast into the initial value
 form using certain special cases of the assumptions made.

 The metric is usually presented as
\be
\label{gends} ds^{2}=P(R)dt^{2}-P(R)^{-1}dR^{2}-R^{2}d\Omega^{2}
\ee
where $P(R)=1-2M/R+E^{2}/R^{2}$
 We perform the following transformation
\bea
 t&=& T+\int \frac{g(R)}{P(R)}dR\\
 r&=&T+\int \frac{1}{g(R)P(R)}dR
\eea
where $g$ is chosen so that $P(R)/(1-g^{2}(R)) > 0$ which  maintains the
 signature of the transformed metric.

The metric reduces to the general spherically symmetric form (\ref{sphline}) with
 \bea
e^{\sigma}&=&\frac{P}{1-g^{2}}\\
e^{\omega}&=&\frac{g^{2}P}{1-g^{2}}
\eea 

One can easily check that this is recovered from the assumptions of the previous section with the particular choice,
$\psi=1$ and $\sigma=Q=\ln{P/(1-g^{2})}$. In order to obtain a solution
for $R$ in terms of $t$ and $r$, we choose $A(r)={\cal A}(r)=1$
with $2m_{o}(r)=2M$  and ${\cal B}(R)=-RP(R)+R-2M$.
Using this in equation (\ref{III}) we obtain
\be
\label{bifur}r-t=\int\frac{1-g(R)^{2}}{P(R)g(R)}dR
\ee

The Reissner Nordstr\"{o}m metric can now be re-cast as

\be
\label{line}ds^{2}= \frac{P(R)}{1-g(R)^{2}}dt^{2}- \frac{g(R)^{2}P(R)}{1-g(R)^{2}}dr^{2}-R^{2}d\Omega^{2}
\ee 

 We can choose the function $g$ to be a constant $g_{o}$ with the condition $1-g_{0}^{2} > 0$ needed ($P>0$ for $R>R_{+}$).   In case $P$ becomes negative (which it does when $E \le M$ between the roots $R_{-}$ and $R_{+}$), we choose $g$ as follows. 
 For the region $R \ge R_{+}+ \epsilon$, we choose $g(R)=g_{0}$. 
 For $ R_{+}+ \epsilon \ge R  \ge R_{+}- \epsilon$ we choose $g$ to be a $C^{\infty}$ function ( $C^{\infty}$ at the endpoints of the interval also). Further
 $g(R_{+})=1$ is chosen with the condition
 $ \lim_{R \rightarrow R_{+}}\frac{P(R)}{1-g^{2}(R)} > 0$. 
For $R_{+}- \epsilon \ge R$, we choose g to be a constant $g_{1}>1$.     

Additionally,
 we need to set $\dot{R} <0$ to mimic a collapse scenario. This needs $g > 0 $. 
  
 	The solution of (\ref{bifur}) can now be obtained explicitly 
($R \ge R_{+} + \epsilon$) .
\be
\label{R}r-t=\frac{1-g_{o}^{2}}{g_{o}}\left[ R+2M/2 ln(R^{2}-2MR+E^{2})+\frac{(2M)^{2}-2E^{2}}{2}\right]H(R)
\ee
 where $H$ is defined as

\bea
H(R)& = & (E^{2}-M^{2})^{-1/2}\left(tan^{-1}(\frac{R-M}{(E^{2}-M^{2})^{1/2}})-tan^{-1}(\frac{-M}{(E^{2}-M^{2})^{1/2}}\right) \nonumber \\
 & &   \mbox{iff $E^{2}>M^{2}$} \\
 H(R)  & = & (M-R)^{-1}R/M  \nonumber \\
 & &  \mbox{iff $E^{2}=M^{2}$} \\ 
 H(R)  & = & 1/2 (M^{2}-E^{2})^{-1/2}
\ln{\frac{1-\frac{R}{M+\sqrt{M^{2}-E^{2}}}}{{1-\frac{R}{M-\sqrt{M^{2}-E^{2}}}}}} \nonumber \\  &  & \mbox{iff $E^{2}<M^{2}$}
\eea

For $R \le R_{+}- \epsilon$ (which would be needed in the sections further), we simply replace $g_{0}$ in (\ref{R}) by $g_{1}$.

For $R <R_{-}$, the case we would need for leading order behaviour of $r-t$ in terms of $R$, we note that $1-g^{2} > 0$ (as P is positive again). It is not 
difficult to see that this behaviour of $r-t$ in terms of $R$ is cubic. One can check explicitly that the coefficients of the linear and quadratic terms in R 
cancel out in the equation ( similar to \ref{R} at the leading order) and one 
obtains the following
\be
\label{Rlo}r-t=\frac{1-g(0)^{2}}{g(0)}KR^{3}
\ee 
where

\bea
K&=&\frac{1}{3E^{2}}   \nonumber\\
 & &   \mbox{iff $E^{2}>M^{2}$} \nonumber \\
K&=& \frac{-1}{3E^{2}}  \nonumber\\
 & &  \mbox{iff $E^{2}=M^{2}$} \nonumber \\
K&=& \frac{M^{2}}{3E^{4}}\left[11-12\frac{M^{2}}{E^{2}}-\frac{3E^{2}}{2M^{2}}\right]  \nonumber\\
  &  & \mbox{iff $E^{2}<M^{2}$}
\eea

Thus $K$ and therefore $r-t$,  is positive only for the case $E^{2}>M^{2}$.
It is this leading order behaviour that is important for further analysis. 
 
The behaviour of the right side of equation (\ref{R}) can be easily checked to be monotonous. Thus, $R=0$ curve is the locus $t-r=0$. It is also easily checked that the Kretschmann scalar diverges when $R=0$.  Thus, the singularity occurs when $t-r=0$. Also, the central singularity forms at $(t=0,r=0)$.

Using the monotonicity and the leading order behaviour (\ref{Rlo}),
we note that in the case $E<M$, the singularity is absent for all $t<0$ at
 $r=0$. We begin with an initial surface $t=t_{in}$. The initial data evolves 
from there to form the singularity. 

Thus we have cast the case $E<M$ in a manner which enables us to compare
 with spherical dust collapse.
 
 The $E>M$ case can also be addressed similarly. In this case, one cannot
 have $t<0$ at all for $r=0$. Thus the surfaces of constant $t$ begin
 from $t=0$ and thus include the singularity from the beginning.

\subsection{Nakedness of Reissner Nordstr\"{o}m Solution}

 At this juncture it is easily possible to demonstrate that the singularity
 is locally naked. It is done in a manner  equivalent to \cite{roots}.
 
Casting the metric into the usual double null form, 
\be
ds^{2}=P(R)dudv+R^{2}d\Omega^{2}
\ee
we find  
$dR/dv$ along $u=constant$ rays (outgoing rays). It is easy to see that 
\be
\frac{dR}{dv}=\frac{P(R)}{2}
\ee
 Evaluating this as $R \rightarrow 0$, we find that it diverges positively independent of $E$ and $M$. Thus Reissner Nordstr\"{o}m singularity is always locally naked.
  
It is useful to cast the initial conditions in terms of $p_{r}$ and $2m$
instead of $E$ and $M$.  
\be
E^{2}=R(t_{in},r)^{4}p_{r}(t_{in},r)
\ee
and
\be
2M=2m(t_{in},r)-R(t_{in},r)^{3}p_{r}(t_{in},r)
\ee
Thus depending on how $2m(t_{in},r)$ the initial mass (at any $r$) compares with
the radial pressure dependent quantity $R(t_{in},r)^{3}p_{r}(t_{in},r)+2R(t_{in},r)^{2}\sqrt{p_{r}(t_{in},r)}$, one has eventually either a naked or a covered singularity.

Further, we examine the nature of initial data for regularity. 
We need to consider the behaviour of $R(t_{in},r)$, for this purpose. We use
equation (\ref{Rlo}) and the monotonicity of the right side of (\ref{R}).
In the case $E\le M$, $R(t_{in},r)$ does not vanish at any $r$ and hence $p_{r}$ as well as $\rho$ do not diverge anywhere on the initial Cauchy surface. The same is 
not true however in the case $E>M$. We know that the singularity is met by all
 Cauchy surfaces as pointed out in the previous section. $R(t_{in},r)$ vanishes
 at $(t=0,r=0)$. Using (\ref{Rlo}) and (\ref{R}), we find that $R(t_{in},r)$ 
vanishes at points on each surface of constant positive $t$. Since the
 surfaces begin from $t=0$ with increasing $t$, $p_{r}$ as well as $\rho$ 
diverge at a point on the surface $t=0$ considered as initial ( and
 on the subsequent surfaces also).

%
%

\begin{figure}[!h]
\parbox[b]{10.99cm}
{
\epsfxsize=10.00cm
\epsfbox{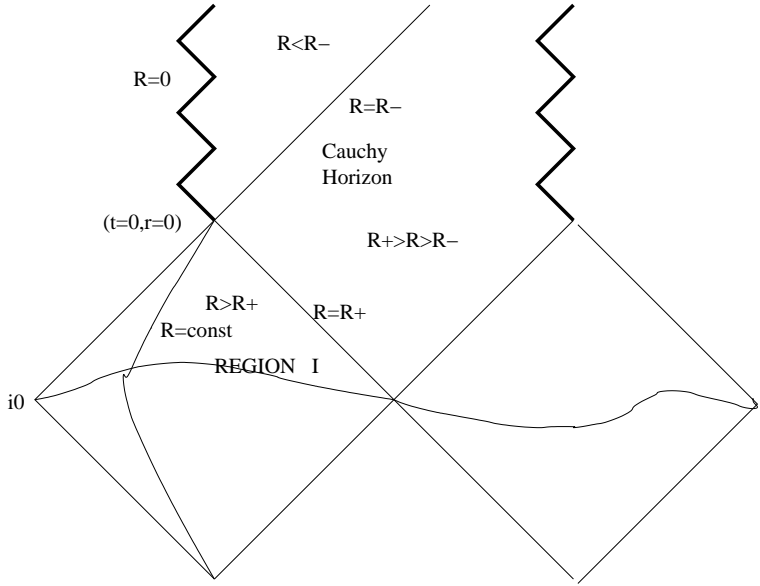}
}
\caption{Penrose diagram for Reissner Nordstr\"{o}m case $E<M$, showing a Cauchysurface across Region I.}
\end{figure}

The singularity develops eventually and a Cauchy horizon forms.

\section{Quantum Stress Tensor for Reissner Nordstr\"{o}m solution}
 We now obtain the quantum stress tensor for the Reissner Nordstr\"{o}m case $E<M$ .
 It should be noted here that we employ the Wald axioms all over the region - when actually, we only demand that one begins with a state of the Hadamard form near the initial Cauchy surface. This supports the suggestion that the
 divergence on the Cauchy horizon we obtain is not an artefact of assumptions on the quantum stress tensor prescription or on the state, made locally. One could certainly work out the same expressions using the standard co-ordinates
 as done by Hiscock, under local validity of Wald axioms. Nevertheless, we use
 the co-ordinate system in the previous section. Thus, we work explicitly with
 co-ordinates covering the Cauchy development and implicitly require the assumption Hadamard property only on the initial Cauchy surface. 

 We begin with the explicit expression of the line element in the (t,r) co-ordinates i.e. in the form (\ref{line}).
\be
ds^{2}= \frac{P(R)}{1-g^{2}(R)}dt^{2}- \frac{g^{2}(R)P(R)}{1-g^{2}(R)}dr^{2}-R^{2}d\Omega^{2}
\ee 

where 
\be
P(R)=1-2M/R+E^{2}/R^{2}
\ee

Since we investigate regions near the Cauchy horizon $R=R_{-}$, we use the
 fact that $g=g_{1}$.
We suppress the angular part and obtain the expressions for the quantum stress 
tensor using the results (\ref{genq}) mentioned earlier.
We cast the resulting 2-d metric in the double null form 
\be
 ds^{2}=C^{2}(u,v)dudv
\ee
where $u=t-g_{1}r$, $v=t+g_{1}r$ and
\be
 C^{2}(u,v)= P(R)/(1-g_{1}^{2})
\ee

The expressions for the quantum stress tensor can be easily worked out using (\ref{genq}). 
\bea \label{actual}
<T_{uu}>&=& -1/12\pi \left(\frac{g_{1}+1/2}{g_{1}^{2}-1}\right)^{2}\left[P'^{2}/4-PP''/2\right] + {\cal AA}(u) \\
<T_{vv}>&=&-1/12\pi \left(\frac{g_{1}-1/2}{g_{1}^{2}-1}\right)^{2}\left[P'^{2}/4-PP''/2\right] + {\cal BB}(v) \\
<T_{uv}>&=&{\cal R}/24\pi
\eea
where ' denotes derivative with respect to R and ${\cal R}$ is the Ricci scalar.

Here, the functions ${\cal AA}$ and ${\cal BB}$ from (\ref{genq}) are chosen to
be finite like in \cite{hiscock}. We are guaranteed that they
 stay finite once we choose them on a spacelike surface, in particular, the
 initial Cauchy surface. This is shown using a simple argument. We present that below.
	
 Since the state is chosen to be Hadamard near the initial Cauchy surface, the 
Wald axioms yield finite expressions for the quantum stress tensor \cite{BD}. 
Then, using equation (\ref{actual}) 
 evaluated near the initial surface, it can be easily checked that the
 functions ${\cal AA}$ and ${\cal BB}$ have to be finite there. The initial Cauchy surface is $t=t_{in}$. Note that $u$ and $v$ depend only on
 $t$ and $r$. Specifying any function of $u$ on this initial surface would
 amount to specifying it completely. Similarly with any function of $v$. 
 This completes the argument.

Returning to equation \ref{actual}, we would need to examine behaviour of $P$ . Real roots of $P(R)=0$ exist when $E\le M$. It should also be noted that the 
metric $C^{2}$ diverges when $R=R_{-}$, the Cauchy horizon. The null co-ordinates need to
 be suitably transformed so as to ensure the regularity of the metric on the
 Cauchy horizon.\footnote{This is similar to the issue of introducing Kruskal
 co-ordinates for the Schwarzschild metric to ensure regularity at the event
horizon. See also \cite{hiscock}, \cite{2d}} We then calculate the stress tensor using these co-ordinates.
     
We perform the following co-ordinate transformation which ensures regularity
of the metric.
\be
W=W_{0}\exp (-\tau u)
\ee
\be
Z=Z_{0}\exp (\tau v)
\ee
where $\tau$ is any positive constant having physical dimension of inverse 
length and $W_{0}$ and $Z_{0}$ are constants of physical dimension length.

One then transforms all tensors to the (W,Z) co-ordinate system.
 
One can check that the metric remains non-zero and finite across the Cauchy horizon. 
\be
g_{WZ}= \frac{R_{+}-R_{-}}{\tau^{2}W_{0}Z_{0}R_{-}^{2}} \exp \left( \frac{2 \tau R_{-}^{2}}{R_{+}-R_{-}}\right) + O(R - R_{-})
\ee
 The $<T_{WW}>$ component can be easily seen to diverge positively, with the 
leading order behaviour $\frac{1}{R^{2}}$. One can also check this explicitly by transforming $<T_{uu}>$ to $<T_{WW}>$. In terms of $W$ the divergence is 
 $\frac{1}{W^{2}}$ to the leading order
 in the approach to the Cauchy horizon. Similar behaviour is reported in the
 spherical dust case \cite{2d}.

\section{Semiclassical Radiation in Reissner Nordstr\"{o}m spacetime}

We now consider the validity of the semiclassical
 approximation in the context of Reissner Nordstr\"{o}m background. 
	If the semiclassical picture were to be followed right upto the
 Cauchy horizon in the Reissner Nordstr\"{o}m background, the results would 
suggest an
 intense particle creation accompanying the naked singularity leading to
 a burst of energy on the Cauchy horizon.  This behaviour is similar to dust as pointed out in the previous section. The semiclassical picture however can be 
 taken to be physically appropriate only upto a limit, beyond which quantum gravity has to 
be invoked. At the same time, backreaction effects must also be considered. We address these issues in the  Reissner Nordstr\"{o}m case and compare with
 dust.
The quantum gravity  limit is worked out by assuming it to be reached when 
curvature scales (of the central region of the dust cloud considered) reach 
the Planck scale \cite{tpplanck}. This happens to occur one Planck length 
before the Cauchy horizon forms in the dust case. The Planck length cutoff is 
crucial for further physical interpretation for the dust case. For instance, if
 the cut off had occured much closer to the Cauchy horizon, the semiclassical
 calculation, which indicates the occurence of a burst, would have been well 
within acceptable range of applicability (assuming that backreaction had 
remained negligible).  

 	This crucial feature is clearly brought out in the Reissner Nordstrom
 case.

  We first show a key feature which concerns boundedness of curvature in the
 $E<M$ case. 
  From equation (\ref{m'}), putting in the assumptions about ${\cal B}$, $A$ and $2m_{0}$, we find that the density is given by   
\be
\label{dens} 8 \pi \rho= \frac{E^{2}}{R^{4}}
\ee
and the radial pressure is given by
\be
\label{press} 8 \pi p_{r}= \frac{E^{2}}{R^{4}}
\ee
 As is expected, both diverge as the central singularity $(t=0,r=0)$ is
 approached. Planck scale values are reached when $t \sim -t_{Planck}$ where
 $t_{Planck}$ is the Planck time. The curvatures grow unboundedly as $R$ decreases to $0$. This would happen only after the curve $R=R_{-}$ is crossed. That 
being the Cauchy horizon, one infers that the curvatures are bounded within 
the Cauchy development. This boundedness within the full Cauchy development is the key feature which makes this example special. We were able to show it explicitly
 by casting the region of interest in suitable co-ordinates enabling easier identification of the full Cauchy development in the problem. 

One important consequence of this is that  curvatures do not reach Planck scales in general before the Cauchy horizon. In 
particular, it is guaranteed if one chooses $R_{-}$ to be much larger than the
 Planck length scale. Thus, we make a distinction between Planck curvature
 limit and Planck length cutoff. The latter is simply a null surface emerging 
out of an event one Planck length before the central singularity forms. We study the semiclassical behaviour upto the Planck length cutoff which we find similar to the dust case.
 
In order to find the total energy $E$ emitted to $\cal{I}^{+}$, one needs
to integrate the power radiated upto the Planck length cutoff. We choose 
$\tau$ and $W_{0}$ such that $g_{WZ}$ tends to unity as $R \rightarrow \infty$.
 
\be
{\cal E} = \int_{\infty}^{t_{Planck}^{2}}P(W)dW 
\ee
where P is the power which turns out to be $<T_{WW}>$. At this stage one
 can work in the full 4-d picture also but needs to resort to geometrical
 optics approximation for calculating the expression for $P$. We describe that
 here.
Following \cite{Ford&Parker} we express the radiated power\footnote{Assuming a conformally coupled field} as
\be
P(W)=\frac{\hbar}{48 \pi}\left(\frac{G^{||}}{G^{|}}\right)^{2}
\ee
where $|$ denotes derivative with respect to the argument. 
$G$ is determined by the centre $r=0$ being cast as the locus
$v=G(W)$. This result is valid for all spherically symmetric backgrounds as the derivation involves modes passing through the centre irrespective of singularity formation at a later stage.

It can be easily checked from this definition that 

\be
G(W)= -\frac{1}{\tau} \ln{W/W_{0}}
\ee

The expression for energy radiated is thus obtained as
\be
{\cal E}= \frac{\hbar}{48 \pi W}
\ee 
 If $W$ is limited to one Planck time before the Cauchy horizon at ${\cal I^{+}}$, then  ${\cal E}$ can be seen to be of the order of the Planck mass.

 This suggests that the data leading to a naked singularity does indeed stay
 largely classical till it reaches very near the Cauchy horizon. Nevertheless,
 a  burst does occur beyond about a Planck length. In order to rule out such a 
 burst on physical grounds, we cannot invoke the Planck curvature limit like in
 the dust case. 

  The issue of backreaction sheds light on this matter and the result is the 
other important consequence of the key feature of boundedness of curvature in the Reissner Nordstrom case. 

We briefly recount the analysis for dust \cite{haradabr}. The energy density
 $<T_{\mu \nu}>u^{\mu}u^{\nu}$ as seen by a co-moving observer moving with 
four velocity $u^{\mu}$, represents the energy density associated with quantum
 fields. The classical background density is then compared with this. The 
background density grows unboundedly as the central singularity is approached.
 Especially near the centre, the background density is larger than the energy 
density of the scalar field, till the central singularity forms.
 This indicates that the backreaction is insignificant during the semiclassical evolution.  

  The situation is different in the Reissner Nordstrom case. The quantum stress
 tensor grows unboundedly while the background density (\ref{dens}) and 
pressure (\ref{press}) remain finite . It is thus impossible for the 
background curvatures to dominate over the quantum field energies at any event
 before the Cauchy horizon.  It therefore turns out that the Reissner 
Nordstrom metric would be unsuitable as a background near the Cauchy horizon 
under considerations of backreaction.

\section{Discussion and Conclusion}

The Schwarzschild solution has been cast in a suitable initial value form 
using co-moving co-ordinates \cite{LL}, wherein pressure does not appear 
 in the source term. Our transformation casts Reissner Nordstr\"{o}m solution 
in a similar form, but can be considered as collapse with non zero 
pressure.

The initial radial pressure is related to $E$ and the initial mass function to $E$ as well as $M$. The case $E<M$ can be
obtained from regular initial data. One may interprete that increasing the 
radial 
pressure as compared to the initial density
 tends to make the singularity globally naked, however the initial data in the
 latter case turns out to be singular. 

Casting the Reissner Nordstrom metric 
 in a particular co-ordinate system has enabled us to give an example where a naked singularity forms but curvatures are bounded within the full Cauchy 
development. By making the Cauchy development explicit in terms of co-ordinates,one can choose to work with certain states like Hadamard states which guarantee
the global validity of Wald axioms, once they are specified near the initial 
 Cauchy surface. This provides stronger support for physical interpretation
 of the results obtained in the Cauchy development, in particular near the
 Cauchy horizon. The results on which these interpretations
 are based could have been drawn using the usual co-ordinates in which
 the Reissner Nordstrom solution is presented (with a suitable modification
 on the Cauchy horizon). However, a separate justification for the global
validity of Wald axioms would have had to be given. We have avoided this.

 The semiclassical analysis recovers Hiscock's result for $E<M$ which shows 
that the quantum stress tensor diverges on the Cauchy horizon, however with 
assumptions now made only on initial data. This indicates that the 
suggestion that there is burst of energy from the singularity is actually a 
consequence of choice of certain initial data. 
	
 In actual physical systems, the burst could be large but finite. To examine 
this further, we analyze the 
domain of semiclassical validity and radiation therein to conclude about the 
behaviour of initial data leading to naked singularities.
Only one Planck mass of the mass of the classical configuration is evaporated 
away upto one Planck time of the classical Cauchy horizon. This behaviour is 
similar to the spherical dust case.
 We thus provide an example of an initial data configuration which collapses 
almost intact throughout the semiclassical domain upto the Planck length cutoff.
 However, in this example, the semiclassical domain  extends beyond this 
cut-off.
 These two results put together support the interpretation that a sudden 
burst occurs near the Cauchy horizon, possibly for some initial data 
configurations. 

 The picture of semiclassical collapse in the dust model has been used to 
emphasize the need to invoke quantum gravity for understanding the spacetime 
region classically corresponding to the Cauchy development, 
in spherical symmetry. In the Reissner Nordstr\"{o}m metric, on the contrary, 
that is not necessary.  This would have been significant physically but for
 the backreaction factor. The latter, 
however, does not clearly indicate the occurence of a burst in the Reissner 
Nordstr\"{o}m case.
 Rather, prominence of backreaction over the quantum energy seen near the
 Cauchy horizon suggests that the Reissner Nordstr\"{o}m case appears to be 
 unsuitable as a background in the region near the Cauchy horizon. 
 We note here that physical unsuitability near the horizon is also 
 suggested by non-local quantities or observer dependent 
divergences (eg. \cite{poisrael}, \cite {hiscockrn}). The analysis was carried out 
considering backreaction of perturbations to the geometry near the horizon.
 However, we note that the unperturbed solution is
 certainly valid as a background over a period of evolution. We recover 
the same result comparing the quantum energies with the curvature.  In fact, 
the solution behaves almost entirely classically throughout the evolution.

	Since we have been emphasizing the region near the Cauchy horizon,
 a few comments on its classical nature are in order. Although the horizon
 will manifest as a surface generated by null rays in spacetime, behaviour
 of certain quantities along geodesics show irregularities. For a precise
 description see \cite{burko}. However, the horizon does not qualify to be
 a physical barrier in the classical sense as the so called 'singularity'
 there is weak in the Tipler sense (volumes stay bounded below in approach to
 the horizon). Thus the semiclassical results which do in fact suggest drastic
 physical behaviour near the horizon, assume significance.
  	
       We conclude that behaviour in spherical collapse, involves the 
interplay of the semiclassical domain, onset of quantum gravity regime and 
prominence of backreaction, which need not be typical. We draw this conclusion 
based on a premise for initial conditions, unlike previous work. It may be 
still true that the contrasting physical behaviour of two kinds of initial data, indicated by  spherical dust, is typical. One possible way for that 
 to hold is the likely prominence of backreaction whenever the onset of 
 quantum gravity is delayed, much like in the example given. It is not known 
if this could just be the case for backgrounds which are physically reasonable. For future work, we therefore suggest that more realistic examples be 
considered.

\section{Acknowledgements}
 Part of this work was carried out by S.B.
 at Albert Einstein Institut, Max Planck Institut f\"{u}r Gravitationsphysik,
 Bundesrepublik Deutschland.
 S.B. wishes to acknowledge his well-wishers, especially his parents, 
 Dr. Madhavi Barve,  Al. Barve (deceased) for continual support.


\begin{thebibliography}{mm}
\bibitem{bojowald} Martin Bojowald {\it Phys.Rev.Lett.} {\bf 86} (2001) 5227
\bibitem{tpplanck} T.P.Singh gr-qc/0012087 (and references therein) to appear
 in conference proceedings of {\it JGRG10} September 2000, Osaka,Japan. In 
particular, see T. Harada, H. Iguchi, K. Nakao, T.P. Singh, T. Tanaka and 
C.Vaz  {\it Phys.Rev. D} {\bf 64} (2001) 041501
\bibitem{hiscock} W. Hiscock {\it Phys. Rev. D} {\bf 15} (1977) 3054; S.A. Fulling {\it Rep. Prog. Phys}{\bf 41} (1978) 1313
\bibitem{hiscockrn} W.A. Hiscock {\it Phys. Lett. A} {\bf 83} (1981) 110
\bibitem{penrose} R. Penrose, in {\it Battelle Rencontres} pg. 222, eds. C.M. DeWitt and J.A. Wheeler), W.A. Benjamin
\bibitem{brady} Patrick R. Brady, to appear in {\it Proceedings of X Brazilian School on Cosmology and Gravitation}
\bibitem{poisrael} E. Poisson and W. Israel {\it Phys. Rev. D} {\bf 41} (1990) 1796 
\bibitem{BD} N.D. Birrel and P.C.W. Davies {\it Quantum Fields in Curved Space} Cambridge University Press
\bibitem{waldqft} R.M.Wald {\it Quantum Field Theory in Curved Spacetime and Black Hole Thermodynamics} University of Chicago Press
Press
\bibitem{hadamard} Wolfgang Junker {\it Rev. Math. Phys.} {\bf 8} (1996) 1091;ibid. (erratum) {\it Rev. Math. Phys.} {\bf 14} (2002) 511; M.J.Radzikowski {\it Commun. Math. Phys.} {\bf 179} (1996) 529
\bibitem{kaywald} B.S.Kay and R.M. Wald {\it Phys. Rep.} {\bf 207} (1991) 49
\bibitem{FSW} S. Fulling, M. Sweeny, R. Wald {\it Commun. Math. Phys.} {\bf 63} (1978) 257; further refinements of the proof are also worked out based on propagation of singularities theorem in microlocal analysis-see M.J. Radzikowski {\it Commun. Math. Phys.} {\bf 179} (1996) 529 ; ibid. {\it Commun. Math. Phys.} {\bf180} (1996) 1
%
%
\bibitem{LL} L.D.Landau and E.M.Lifshitz {\it Classical Theory of Fields} Pergamon Press
\bibitem {tang} S. Barve T.P. Singh and Louis Witten {\it Gen. Rel. Grav.} {\bf 32} (2000) 697
\bibitem{roots} T.P. Singh and P.S. Joshi{\it Class. Quant. Grav.}{\bf 13} (1996) 559
\bibitem{2d}S. Barve T.P. Singh C. Vaz and Louis Witten{\it Phys. Rev. D} {\bf 58} (1998) 104018.
\bibitem{Ford&Parker} L.H. Ford and L. Parker {\it Phys. Rev. D} {\bf 17} (1978) 1485
\bibitem{haradabr} Hideo Iguchi and Tomohiro Harada {\it Class. Quant. Grav.}{\bf 18 } (2001) 3681
\bibitem{burko} Lior M. Burko and Amos Ori {\it Phys. Rev. D} {\bf 56} (1997) 7820 (also includes a detailed exposition of metric equations using source terms
 for a scalar field and electromagnetic radiation); L.M. Burko {\it Phys. Rev. Lett.} {\bf 79} (1997) 4958; ibid. {\it Phys. Rev. D}{\bf 60} (1999) 104033
\end{thebibliography}
\end{document}